\documentclass{WileyMSP-template}
\usepackage{multirow}%
\usepackage{amsmath,amssymb,amsfonts}%
\usepackage{amsthm}%
\usepackage{mathrsfs}%
\usepackage[title]{appendix}%
\usepackage{xcolor}%
\usepackage{textcomp}%
\usepackage{manyfoot}%
\usepackage{booktabs}%
\usepackage{algorithm}%
\usepackage{algorithmicx}%
\usepackage{algpseudocode}%    
\usepackage{listings}%
\usepackage[pdftex,colorlinks=true,bookmarks=false,citecolor=blue,urlcolor=blue]{hyperref} %pdflatex
\hypersetup{pdfstartview={FitH},pdfpagemode={UseNone}, breaklinks=true,
            colorlinks,linkcolor=blue, citecolor=blue, urlcolor=blue,
             bookmarksopen=true, pdfnewwindow=true}
\usepackage[all]{hypcap}

\usepackage{parskip}
\graphicspath{{pict/}{figures/}{}}

\usepackage[english]{babel}
\usepackage{ragged2e}

\begin{document}
\justifying
\sloppy

% \pagestyle{fancy}
% \rhead{\includegraphics[width=2.5cm]{vch-logo.png}}

% \title{Flat-optics quantum light sources: fundamentals and applications}
\title{Engineering Quantum Light Sources with Flat Optics}

\maketitle

% Author: Please give full first and last names for authors and include * after the name of all corresponding authors

\author{Jinyong Ma$^{\dagger,*}$,}
\author{Jihua Zhang$^{\dagger,*}$,}
\author{Jake Horder$^{\dagger}$,}
\author{Andrey A. Sukhorukov,}
\author{Milos Toth,}
\author{Dragomir~N.~Neshev,} 
\author{Igor Aharonovich$^{*}$} 
\\$^{\dagger}$Co-first authors with equal contribution

% \author{}

% % Dedication

% % \dedication{de}

% Affiliations: Please provide adacemic titles (Prof. or Dr.) for all authors where applicable, and include an institutional email address for all corresponding authors
\begin{affiliations}

Jinyong Ma, Jihua Zhang, Andrey A. Sukhorukov, and Dragomir N. Neshev \\
ARC Centre of Excellence for Transformative Meta-Optical Systems (TMOS), Department of Electronic Materials Engineering, Australian National University, Canberra, 2600, Australia\\
Email: jinyong.ma@anu.edu.au

Jihua Zhang \\
Songshan Lake Materials Laboratory, Dongguan, Guangdong 523808, P. R. China \\
Email: zhangjihua@sslab.org.cn \\

Jake Horder, Milos Toth, and Igor Aharonovich \\
ARC Centre of Excellence for Transformative Meta-Optical Systems (TMOS), School of Mathematical and Physical Sciences, University of Technology Sydney, Australia \\
Email: igor.aharonovich@uts.edu.au
% Email Address:

\end{affiliations}

% Keywords: Please provide a minimum of three and a maximum of seven keywords, separated by commas

\keywords{Quantum light source, photon pairs, single photon source, flat optics, metasurface}

% Abstract should be written in the present tense and impersonal style (i.e., avoid we), and be at most 200 words long
\begin{abstract}
Quantum light sources are essential building blocks for many quantum technologies, enabling secure communication, powerful computing, precise sensing and imaging. Recent advancements have witnessed a significant shift towards the utilization of ``flat" optics with thickness at subwavelength scales for the development of quantum light sources. This approach offers notable advantages over conventional bulky counterparts, including compactness, scalability, and improved efficiency, along with added functionalities. This review focuses on the recent advances in leveraging flat optics to generate quantum light sources. Specifically, we explore the generation of entangled photon pairs through spontaneous parametric down-conversion in nonlinear metasurfaces, as well as single photon emission from quantum emitters including quantum dots and color centers in 3D and 2D materials. The review covers theoretical principles, fabrication techniques, and properties of these sources, with particular emphasis on the enhanced generation and engineering of quantum light sources using optical resonances supported by nanostructures. We discuss the diverse application range of these sources and highlight the current challenges and perspectives in the field.

\end{abstract}

% Text: Please use section headings and subheadings as specified below. For communications, all section headings apart from Experimental Section should be removed
% Please make the first reference to a display item bold: \textbf{Figure 1}
% Do not abbreviate Figure, Equation, etc.; display items are always singular, i.e., Figure 1 and 2.
% Equations are always singular, i.e., Equation 1 and 2, and should be inserted using the {equation} environment, not as graphics
% Please do not use footnotes in the text, additional information can be added to the Reference list.

\section{Introduction}
Flat optics, which are surfaces at sub-wavelength scales, have been the subject of extensive scientific inquiry for over a decade~\cite{Yu:2014-139:NMAT, Shastri:2023-36:NPHOT}. Metasurfaces incorporating nano-patterned structures are one of the pivotal subjects of flat optics, showing ongoing success in miniaturizing bulky optical devices with added functionalities in tailoring the directionality, polarization, wavefront and spectral properties of light~\cite{Neshev:2023-26:NPHOT}. Flat optics was recently introduced to advanced quantum technologies~\cite{Kan:2023-2202759:ADOM}, where metasurfaces have been leveraged to generate, manipulate and detect quantum light sources, offering unlimited potential to realize ultra-compact quantum devices with flexible and customized features~\cite{Wang:2022-38:PT, Sharapova:2023-2200408:LPR}. Here this review focuses on the latest achievements in the metasurface-based quantum light sources.

Quantum light sources are the fundamental component of photonic quantum technologies. We divide the sources into two major categories: (i) probabilistic entangled photon pairs (called signal and idler photons) generated from nonlinear optical interactions such as spontaneous four-wave mixing and spontaneous parametric down-conversion (SPDC), and (ii) on demand, single photons emitted from atom-like transitions in solid-state materials – including quantum dots (QDs) and color centers. The generation of photon pairs adheres to the conservation of energy and momentum, resulting in correlated or entangled photons in polarization, spectral, and spatial degrees of freedom. The non-classical correlation or entanglement features of photon pairs, which do not exist in classical light, empower quantum applications in communication~\cite{Liao:2017-43:NAT, Ren:2017-70:NAT}, sensing, and ghost imaging~\cite{Padgett:2017-20160233:PTRSA, Moreau:2019-367:NRP}. Millimetre- or centimetre-scale nonlinear crystals are conventionally used to produce photon pairs. However, these crystals are not only bulky but manifest limited flexibility in engineering the photon properties. With regards to the other type of quantum light sources, single photon emitters (SPEs) are generally compact and have been discovered in many different platforms, including QDs~\cite{senellart2017HighPerformance}, color centers~\cite{bradac2019QuantumNanophotonics}, and 2D materials~\cite{turunen2022QuantumPhotonics}. While the emitters have found applications in quantum sensing~\cite{Pirandola:2018-724:NPHOT}, secure communications~\cite{Couteau:2023-326:NRP}, and quantum computing~\cite{OBrien:2007-1567:SCI}, there are still open challenges in optimizing their efficiency, purity, integration into practical devices and achieving operation at room temperature~\cite{Azzam:2021-240502:APL}. With continuing advances in nanofabrication technology, it has been a critical task to develop both types of quantum light sources that can address their respective challenges to suit the needs of advanced quantum technologies.

\begin{figure}
  \includegraphics[width=\linewidth]{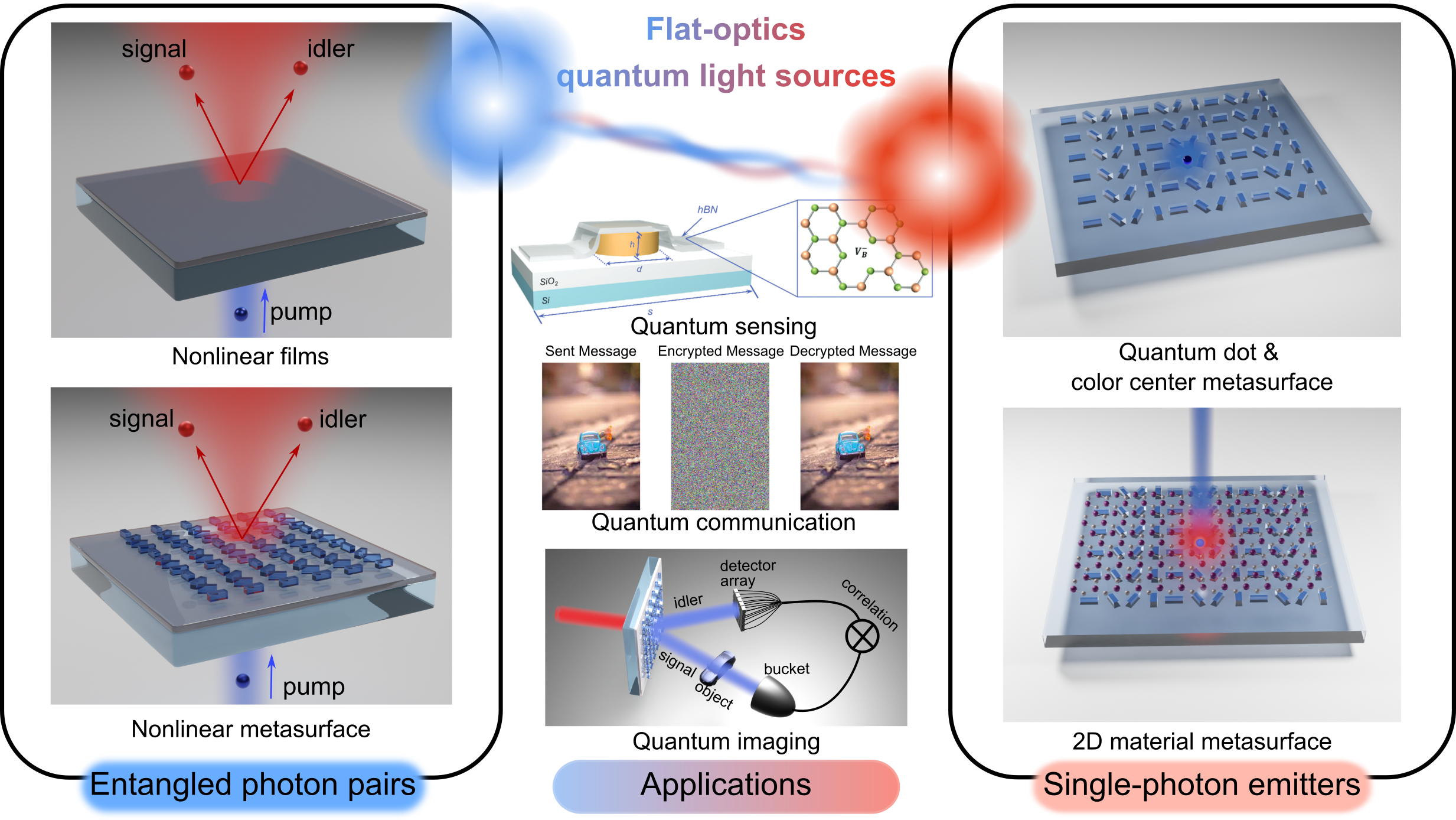}
  \caption{\textbf{Sketch of the review.} We review two types of flat-optics quantum light sources: (i) entangled photon pairs from nonlinear films and metasurfaces (left column) and (ii) single photon emission from an unstructured 2D material, a 2D material metasurface, and an epitaxial quantum dot (QD) metasurface (right column). The sources empower a variety of quantum applications (middle column), including quantum sensing~\cite{yangSpin2022}, quantum communication~\cite{ al2023quantum} and quantum imaging~\cite{Ren23imaging}.}
  \label{fig:sketch}
\end{figure}

Here we will discuss the state-of-the-art advancements in applying flat optics to overcome challenges in quantum light sources. We first introduce functionalities of metasurfaces supporting local and nonlocal resonances, including the enhanced light-matter interaction and multifunctional control of wavefront, polarization, and spectral properties of light. Leveraging these unique features of metasurfaces, we review, as shown in Fig.~\ref{fig:sketch}, the latest development of flat-optics photon pair sources, from ultra-thin nonlinear films to nonlinear optical metasurfaces, which facilitate the enhanced generation of spatially, spectrally, and polarization-entangled photons. Further, we discuss the progress in metasurface-assisted single photon emission from epitaxial QDs and defects in 2D materials, allowing Purcell enhancement, directional emission, and other functionalities. Lastly, we compare the two types of quantum light sources, discuss the prevailing challenges within the field, and offer insights into the potential directions for the future advancement of quantum light sources.

\section{Multi-functional metasurfaces}
\begin{figure}
  \includegraphics[width=\linewidth]{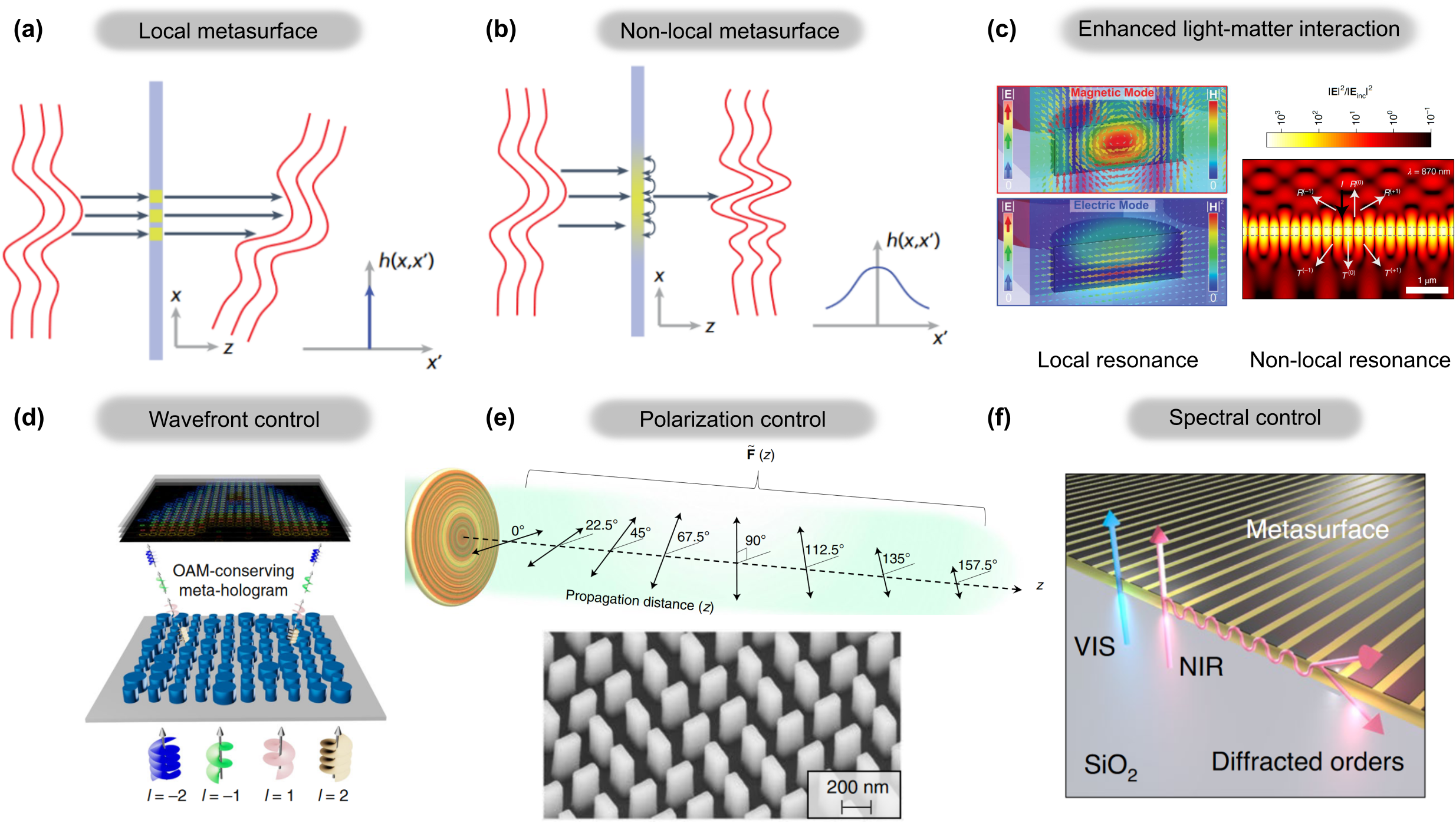}
  \caption{\textbf{Functionalities of metasurfaces.} a)~Local metasurface~\cite{Shastri:2023-36:NPHOT}. The amplitude and phase of light can be manipulated in real space at each individual meta-atoms of local metasurfaces. b)~Non-local metasurface~\cite{Shastri:2023-36:NPHOT}. The function $h(x,x')$ represents the metasurface response in real space. A collective
interaction of light involving many meta-atoms is manifested in non-local metasurfaces. c)~Enhanced light-matter interaction~\cite{Decker:2015-813:ADOM, Song:2021-1224:NNANO}. Optical resonances supported by metasurfaces can enhance light-matter interactions.
d)~Wavefront control~\cite{Diaz-Rubio:2017-e1602714:SCA}. The wavefront of light can be shaped by metasurfaces incorporating spatially-configured arrays of meta-atoms. e)~Polarization control~\cite{dorrahMetasurface2021}.  Metasurfaces facilitate the manipulation of polarization in multiple planes along the direction of light propagation. f)~Spectral control~\cite{Song:2021-1224:NNANO}. Non-local metasurfaces enable fully decoupled optical functions at multiple wavelengths of light. 
 Reproduced with permission. Copyright Year, Publisher. }

%Metasurfaces with arranged birefringent wave plate elements offer the capability to manipulate polarization states of light.
  \label{fig:metasurface}
\end{figure}

Before delving into the flat-optics quantum light sources, we will give a brief introduction to metasurfaces and their functionalities (see Fig.~\ref{fig:metasurface}), which will bridge the gaps between traditional quantum light sources and metasurfaces. 

Metasurfaces, comprising dielectric or metallic nanostructures (known as ``meta-atoms") at subwavelength thickness, serve as excellent platforms for miniaturizing optical devices and providing additional functionalities. The nanostructures can be tailored to control almost all degrees of freedom of incident light, including its polarization, spatial, and spectral properties. Local and nonlocal metasurfaces represent two different categories in metasurface design, each offering unique capabilities in manipulating optical waves. As illustrated in Fig.~\ref{fig:metasurface}(a)-(b), local metasurfaces can independently control the amplitude and phase of light at individual meta-atoms in real space while nonlocal metasurfaces exhibit a collective interaction of light across many meta-atoms. Consequently, these two resonance types manifest distinct angular dispersions and offer different functionalities for applications. 

One key function of metasurfaces is to enhance light-matter interactions, where the optical intensity enhancement is characterized by quality factors (Q factors) of the resonant modes. Figure~\ref{fig:metasurface}(c) depicts the mode profiles for local and nonlocal resonances supported by metasurfaces. The optical modes in local resonances are confined within individual meta-atoms and are defined by their geometries, featuring reasonable Q factors and intensity enhancement at the scale of a few tens. In addition to Mie-type resonance~\cite{Decker:2015-813:ADOM, linPhotoluminescence2022}, local resonances include waveguide-like modes that are supported by nanopillars with different geometries~\cite{chenBroadband2018,linAchromatic2019, wangBroadband2018} and toroidal modes which are distinct from the magnetic and electric dipoles~\cite{wuOptical2018, hassanfiroozi2021toroidal}. These resonances enable the tailoring of Q factors and dispersion. In contrast, optical fields in nonlocal resonances propagate transversely between meta-atoms and are influenced by their couplings. In general, a large lattice of meta-atoms is required to guide the mode propagation. The nonlocal resonances are controlled by the lattice arrangement along with {meta-atom geometries~\cite{bin-alamUltrahighQ2021, Overvig:2022-2100633:LPR}. Recently metasurfaces formed by meta-atom lattices with broken in-plane symmetry were observed to support high-Q resonances, resulting from the distortion of symmetry-protected bound states in the continuum (BIC)~\cite{Koshelev:2018-193903:PRL}. The concept of BIC has opened the door to achieving Q factors $>10^4$~\cite{kang2023applications, Liu:2019-253901:PRL}. 

Metasurfaces can impose desired amplitude and phase changes on incident light, allowing precise control of the light wavefront. This feature was typically realized with localized and low-Q modes of nanostructures~\cite{Yu:2014-139:NMAT, Genevet:2017-139:OPT}, facilitating optical wavefront manipulation for applications like lensing and holography. As illustrated in Fig.~\ref{fig:metasurface}(d), orbital angular momentum (OAM) holography was demonstrated with a metasurface incorporating GaN nanopillars with discrete spatial frequency distributions, enabling lensless reconstruction of OAM-dependent holographic images. Recent attention to wavefront manipulation has shifted towards nonlocal metasurfaces. Díaz-Rubio et al. introduced a novel approach to designing perfect reflectors through energy surface channelling, experimentally validating a highly reflective surface with rectangular metal patches and a metallic plate~\cite{Diaz-Rubio:2017-e1602714:SCA}. Malek et al. demonstrated a versatile multispectral wavefront shaping platform, achieved by employing a stack of metasurfaces, each capable of independently controlling quasi-BIC to tailor optical wavefronts at multiple wavelengths~\cite{Malek:2022-246:LSA}. These works with nonlocal resonances enable wavefront transformation with much higher efficiency than is possible with local metasurfaces~\cite{Shastri:2023-36:NPHOT}. 

An extended application of wavefront control with metasurfaces is the pixelated polarization manipulation of the incident light at the subwavelength scale. Conventionally polarization control is realized with local metasurfaces, where subwavelength anisotropic structures enable substantial refractive index contrast for both TE and TM polarized light~\cite{Hu:2020-3755:NANP}. This allows each meta-atom to induce dual independent phase shifts in transmitted light. Mueller et al. introduced a method to independently impose arbitrary phase profiles on orthogonal states of polarization (including linear, circular, or elliptical) using linear birefringent metasurfaces composed of simple wave plate elements~\cite{Mueller:2017-113901:PRL}. Later, Dorrah et al. have advanced polarization control over multiple planes along the direction of light propagation~\cite{dorrahMetasurface2021}. Recently nonlocal resonances were also developed to control~\cite{Overvig:2020-17402:PRL} and detect~\cite{Hong:2023-134:OPT} polarizations of light at multiple wavelengths. The advanced control over polarization using either local or nonlocal metasurfaces opens up a myriad of applications, including polarization imaging~\cite{arbabiFullStokes2018}, data encryption, display, and optical communication.

Additionally, nonlocal metasurfaces provide a higher degree of spectral control over the light wavefront, as shown in Fig.~\ref{fig:metasurface}(f), which is desirable in many emerging sensing, imaging, holography, augmented reality and nonlinear optics applications. While conventional local metasurfaces can be tailored to possess specific transfer functions at discrete angles and frequencies, their optical response exhibits only limited dependence on the angle and exhibits relatively low Q factor~\cite{Shastri:2023-36:NPHOT}. Nonlocal metasurfaces provide an excellent solution to this challenge by virtue of their pronounced spatial and spectral dispersion, as well as potentially high Q factor. Recent studies have leveraged these platforms to effectively control light with independent functions across various wavelength bands~\cite{Song:2021-1224:NNANO}. 

These functionalities of metasurfaces serve as the building blocks for engineering quantum light sources using flat optics. They can be used to tailor the quantum states of photons in all degrees of freedom, achieving applications beyond what is possible with traditional means.

\section{Generation of quantum entangled photon pairs from nonlinear flat-optics}
\begin{figure}
  \includegraphics[width=\linewidth]{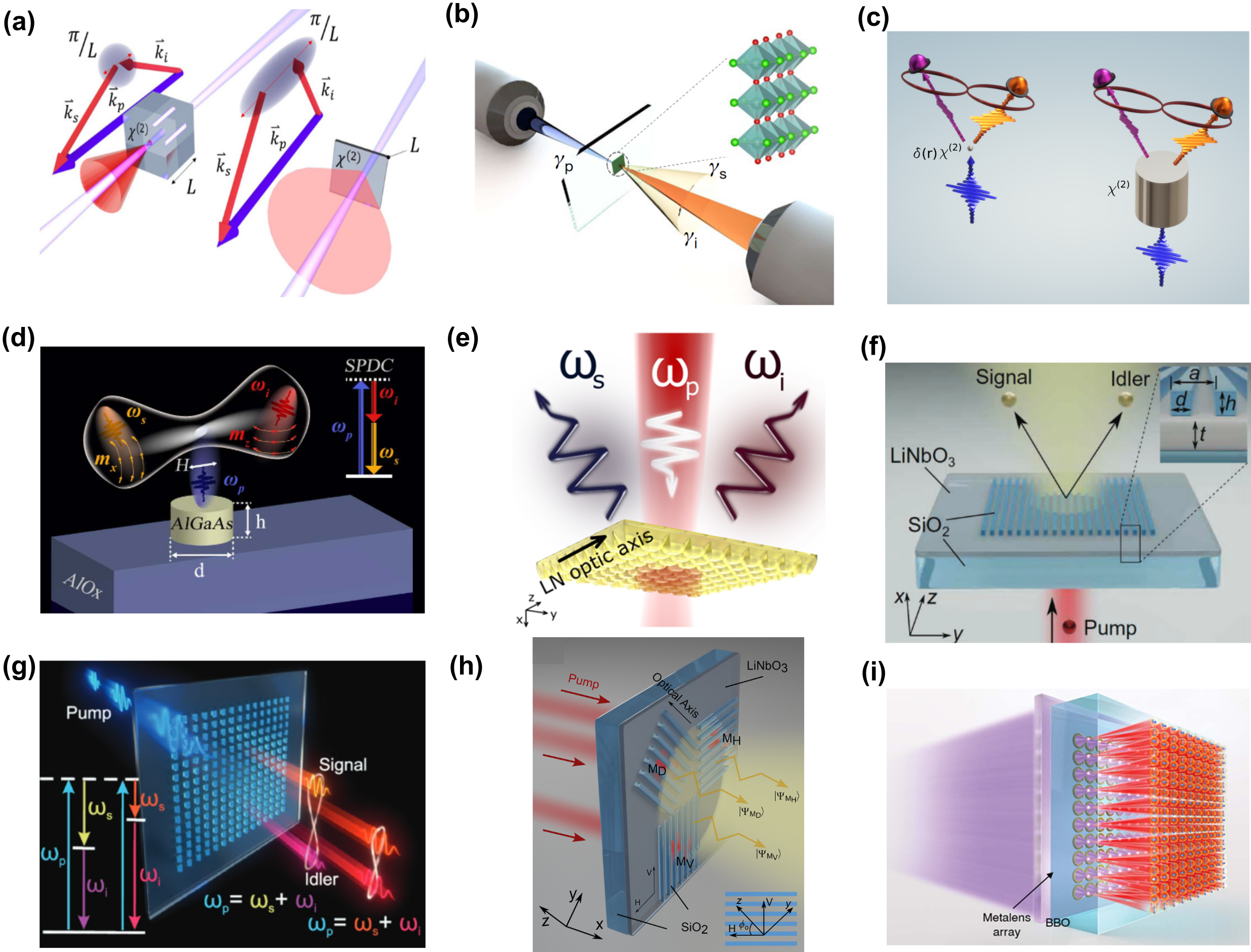}
   \caption{\textbf{Generation of quantum entangled photon pairs from nonlinear flat-optics.} 
% (a,b,c) Photon pairs from nonlinear thin films. (d,e,f) Enhanced generation of photon pairs from nonlinear metasurfaces. (g,h,i) Spatial, spectral, and polarization control of entangled photon pairs with metasurfaces.
% Direction and polarization control. b)~Photon‐Pair Generation with a 100 nm Thick Carbon Nanotube Film~\cite{Lee:2017-1605978:ADM}.
   a)~Photon pairs with very wide spatial and frequency spectra were generated from a lithium niobate film of 6 \textmu m thickness via SPDC, in which the phase matching condition becomes irrelevant~\cite{Okoth:2019-263602:PRL}. 
   b)~Entangled photon pairs were produced from a NbOCl$_2$ crystal, a vander Waals two-dimensional material~\cite{Guo:2023-53:NAT}. 
   c)~Nonlinear Nanoresonators were theoretically predicted as natural sources of Bell states with a large range of emission directions and frequencies~\cite{weissflogNonlinear2024}. 
   d)~A dielectric nanoantenna made of crystalline AlGaAs nanocylinder was used to experimentally demonstrate enhanced SPDC~\cite{Marino:2019-1416:OPT}. 
   e)~The emission rate of photon pairs from resonant metasurfaces was enhanced by two orders of magnitude as compared to unstructured film~\cite{Santiago-Cruz:2021-4423:NANL}. 
   f)~Spatially entangled photon pairs were experimentally generated from lithium niobate nonlocal metasurfaces~\cite{Zhang:2022-eabq4240:SCA}. g)~Resonant metasurfaces were employed to produce complex quantum states in spectral degree of freedom of photons~\cite{Santiago-Cruz:2022-991:SCI}. h)~A lithium niobate nonlinear metasurface incorporating multiplex metagratings enabled the polarization engineering of entangled photons~\cite{Ma:2023-8091:NANL}. i)~High-dimensional and multiphoton quantum source with a linear metasurface. A metalens array is integrated with a bulky nonlinear material to achieve high-dimensional path entanglement of two photons and generation of four and six photons.~\cite{liMetalensarray2020b}. Reproduced with permission. Copyright Year, Publisher.
%d)~It was theoretically proposed that nonlinear metasurfaces supporting BIC resonances can significantly enhance the SPDC process~\cite{Parry:2021-55001:ADP}.
%   Other references to be used in the main text:
% \cite{Poddubny:2016-123901:PRL} https://doi.org/10.1103/PhysRevLett.117.123901
% \cite{Okoth:2020-11801:PRA} https://doi.org/10.1103/PhysRevA.101.011801
% \cite{Jin:2021-19903:NASC} https://doi.org/10.1039/d1nr05379e
% \cite{Mazzanti:2022-35006:NJP} https://doi.org/10.1088/1367-2630/ac599e
% \cite{Sultanov:2022-3872:OL} https://doi.org/10.1364/OL.458133
% \cite{Zhang:2023-10005:COL} https://doi.org/10.3788/COL202321.010005
% \cite{Son:2023-2567:NASC} https://doi.org/10.1039/d2nr05499j
% \cite{Duong:2022-3696:OME} https://doi.org/10.1364/OME.462981
% \cite{Saerens:2023-3245:NANL} https://doi.org/10.1021/acs.nanolett.3c00026
% https://arxiv.org/abs/2311.16036
% https://doi.org/10.1103/PhysRevResearch.5.043228
% https://doi.org/10.1021/acsphotonics.3c01169
}
  \label{fig:SPDC}
\end{figure}

Photon pairs can be generated through the nonlinear processes such as SPDC~\cite{Klyshko:1988:PhotonsNonlinear} or spontaneous four-wave mixing~\cite{HaoyangProgress2024}. These sources have advanced to achieve production rates exceeding millions of heralded single photons per second, allow the preparation of specific quantum states, and exhibit customized spectral characteristics. Here this review focuses on the SPDC process in which a pump photon is split into two photons called ``signal"  and ``idler" via second-order nonlinear interactions. This process adheres to energy conservation, ensuring that their frequencies satisfy $\omega_{p} = \omega_{s}+\omega_{i}$ (with subscripts ``p, s, i" denoting pump, signal, and idler respectively). The SPDC process in nonlinear bulky crystals must also satisfy the phase-matching condition, where the pump wavevector ($k_{p}$) equals the sum of the signal ($k_{\rm s}$) and idler wavevectors ($k_{i}$), i.e., ($k_{p} = k_{ s}+k_{i}$). Transverse phase matching is generally determined by the pump wavevector distribution, while longitudinal phase mismatching, i.e., $\Delta k_l = k_{p,l} - (k_{s,l}+k_{i,l})\neq 0$, is inherently unavoidable due to material dispersion. Several experimental techniques have been introduced to address this issue, such as perfect phase matching by orienting the crystal angles or quasi-phase-matching through periodic poling. These techniques necessitate dedicated temperature control of nonlinear crystals to counteract changes in material dispersion. Only a few nonlinear materials, such as barium borate~\cite{Kwiat:1995-4337:PRL}, periodically poled potassium titanyl phosphate~\cite{jin2014pulsed}, and periodically poled lithium niobate~\cite{Tanzilli:2002-155:EPD}, were developed to achieve the phase-matched SPDC process.

The longitudinal phase matching becomes unimportant as the nonlinear crystal length $L$ decreases to the scale at which the photon pairs are always generated in phase no matter how large $\Delta k_l$ is~\cite{Okoth:2019-263602:PRL, Okoth:2020-11801:PRA, sultanovTemporally2024a, mekhael2023phase}. In other words, $\Delta k_l L \approx 0$ is always satisfied. As shown in Fig.~\ref{fig:SPDC}(a), Okoth et al. report the observation of photon-pair generation free of phase matching~\cite{Okoth:2019-263602:PRL} using a 6 \textmu m thick layer of lithium niobate. The ultrathin structure results in a significantly wider frequency spectrum compared to phase-matched SPDC. While the generation of nanoscale photon pair source was first demonstrated in a 100 nm thick carbon nanotube film~\cite{Lee:2017-1605978:ADM} via four-wave mixing, the generation of entangled photons via SPDC at subwavelength scale was realized in lithium niobate and gallium phosphide nanofilms~\cite{Santiago-Cruz:2021-653:OL, Sultanov:2022-3872:OL},  bottom-up grown lithium niobate microcubes~\cite{Duong:2022-3696:OME} and GaAs Nanowires \cite{Saerens:2023-3245:NANL}. These works bridged the gaps between flat optics and quantum optics, opening the door towards the development of quantum nonlinear metasurfaces. Just recently, Sultanov et al. demonstrated the production of photon pairs in an ultrathin GaP layer with controllable polarization entanglement, by adjusting the pump polarization while maintaining state purity~\cite{Sultanov:2022-3872:OL}. Notably, this field is moving towards two-dimensional materials. Guo et al.~\cite{Guo:2023-53:NAT} introduced niobium oxide dichloride (NbOCl$_2$) (Fig.~\ref{fig:SPDC}(b)), a van der Waals crystal with minimal interlayer electronic coupling, enabling monolayer-like excitonic behavior and exceptionally strong second-harmonic generation. This property enables the first demonstration of correlated photon pair generation through spontaneous parametric down-conversion in ultra-thin flakes. Most recently, Weissflog et al. reported polarization entanglement from a 3R-stacked transition metal dichalcogenide crystal~\cite{weissflogTunable2023a}, where the entanglement is tunable via pump polarization. These works offer the potential for compact on-chip photon sources and high-performance photon modulators in quantum technologies.

Although the photon pair sources based on nonlinear films opened a novel avenue in quantum state generation at the nanoscale, it came at the cost of decreased SPDC efficiency because of ultra-short nonlinear interaction length. Nonlinear dielectric metasurfaces, employing ultra-thin nonlinear materials at the scale of a few hundred nanometers, are emerging as promising platforms for enhanced, compact, versatile, and highly flexible photon pair sources. 

Theoretically, it was proposed that nonlinear nanoresonators supporting local resonances can produce polarization-entangled photons over a very large range of emission directions and frequencies~\cite{weissflogNonlinear2024, Zhang:2023-10005:COL}, as depicted in Fig.~\ref{fig:SPDC}(c). Green's function method~\cite{Poddubny:2016-123901:PRL, krsticNonperturbative2023} along with quasi-normal mode theory~\cite{lalanne2018light, yan2018rigorous, sauvan2021quasinormal} was introduced to analyze the SPDC processes in the nanoresonators, offering a useful SPDC toolbox at the nanoscale. Furthermore, quantum-classical correspondence~\cite{Lenzini:2018-17143:LSA, Poddubny:2020-147:QuantumNonlinear} allows a simple calculation of the two-photon wave function in SPDC through the simulations of its classical counterpart of sum-frequency generation. It was then proposed that entangled photon pairs can be produced from plasmonic metasurfaces~\cite{Jin:2021-19903:NASC} and nonlinear hyperbolic metamaterials~\cite{Poddubny:2016-123901:PRL}. Further enhancement of photon-pair rate was made possible with extended BIC resonances~\cite{Parry:2021-55001:ADP, Mazzanti:2022-35006:NJP}. Parry et al.~\cite{Parry:2021-55001:ADP} theoretically predict a regime of nondegenerate photon-pair generation in nonlinear metasurfaces, driven by the interaction of multiple BIC resonances and enabled by the hyperbolic topology of transverse phase matching, offering the potential for significant enhancements in photon rate and spectral brightness compared to the degenerate regime. Enhanced generation of angle-correlated photon pairs was also predicted in a nonlocal metasurface composed of nanostructured nonlinear metagratings~\cite{Mazzanti:2022-35006:NJP}. 
%These theoretical frameworks established the fundamentals of experimental demonstration. Spectral engineering is impossible with local resonances.

The first experimental demonstration of enhanced photon pair generation in nanoresonators was realized in a dielectric nanoantenna made of crystalline AlGaAs nanocylinder by Marino et al.~\cite{Marino:2019-1416:OPT}, as shown in Fig.~\ref{fig:SPDC}(d). The experiments pave the path for coherent multiphoton quantum state generation through antenna multiplexing. Santiago-Cruz~\cite{Santiago-Cruz:2021-4423:NANL} reported the enhanced photon pair generation in lithium niobate metasurfaces featuring electric and magnetic local resonances~\cite{Fedotova:2020-8608:NANL} across diverse wavelengths (Fig.~\ref{fig:SPDC}(e)). The spectral brightness is significantly increased by two orders of magnitude in contrast to unpatterned films of similar thickness and material, opening the door to flat optics sources of entangled photons. 

%{\color{gray} These works based on local metasurfaces provide limited enhancement in photon-pair rate and present restricted tunability in spectral and spatial modes of optical fields due to the low Q factor and local features of the optical resonances.}

Beyond enhancement, metasurfaces facilitate the generation of complex quantum states in spatial, spectral and polarization degrees of freedom. Zhang et al. predicted and experimentally demonstrated, as depicted in Fig.~\ref{fig:SPDC}(f), spatially entangled photon pair generation using a metasurface with a lithium niobate film and silica meta-grating~\cite{Zhang:2022-eabq4240:SCA}. The spatial correlations of photon pairs violate the Cauchy–Schwarz inequality, indicating multimode entanglement. Photon-pair rates increase 450-fold due to high-quality resonances, offering the potential for miniaturizing quantum devices with ultrathin metasurfaces as room-temperature entangled photon sources. At the same time (Fig.~\ref{fig:SPDC}(g)), Santiago-Cruz et al. produced entangled photons using semiconductor metasurfaces with high-quality factor quasi-bound states in the continuum (qBIC) resonances, enhancing the quantum vacuum field and boosting the emission of nondegenerate entangled photons across multiple narrow resonance bands~\cite{Santiago-Cruz:2022-991:SCI}. Single or multiple resonances, pumped at various wavelengths, can be used to create multifrequency quantum states such as cluster states. A metasurface carrying qBIC also allows the bi-directional emission of photon pairs~\cite{Son:2023-2567:NASC}, which can be used for demultiplexing of photons at the nanoscale. Additionally, Ma et al. introduce a versatile method for engineering polarization states in photon pairs generated from a nonlinear metasurface~\cite{Ma:2023-8091:NANL} (Fig. \ref{fig:SPDC}(h)). It enables the preparation of arbitrary polarization Bell states and qutrits, offering the potential for miniaturized, room-temperature polarization-entangled photon sources in optically re-configurable quantum devices.

It is noted that linear metasurfaces were recently introduced to transform and measure the quantum states of bright photon sources from bulky nonlinear material~\cite{wangQuantum2018a, georgiMetasurface2019a, liMetalensarray2020b, stavQuantum2018e, gao2022multichannel,  li2021NonunitaryMetasurface}. Furthermore, as shown in Fig.~\ref{fig:SPDC}(i), a metalens array integrated with a bulky nonlinear crystal has been employed to generate high-dimensional and multiphoton quantum states~\cite{liMetalensarray2020b}. However, the operation and dimensions of the whole device are still subject to the longitudinal phase-matching and size of the nonlinear crystals and related components. A prospective direction may involve incorporating nonlinear with linear metasurfaces to achieve highly efficient and ultra-compact multi-functional photon pair sources.

\section{Flat optics engineered solid-state SPEs}
\subsection{Quantum dots and color centres}
% As discussed before, 
% Photonic nanostructures can improve the performance of single-photon emitters in several aspects. First, by putting a single-photon emitter into a photonic cavity or resonator, the photon radiative rate can be increased through the Purcell effect, or a reduced pump power is needed to reach a similar photon emission rate. Second, photonic structures can route the emitted photons into a desired free space mode and increase the collection efficiency of the photons into an objective or fiber. Third, photonic structures can control the spectrum, polarization, direction, and OAM of the photons. These controls can increase the degree of indistinguishability of photons from different trials. For epitaxial semiconductor quantum dots, it is possible to generate entangled photon pairs through the biexciton-exciton (XX-X) cascaded radiative process. In this case, a broadband photonic structure can increase the emission rate of both photons with close wavelengths and fidelity of entanglement. 

Solid-state SPEs such as epitaxial semiconductor QDs and color centers in wide bandgap materials are able to generate deterministic single photons. Photonic nanostructures can improve the performance of SPEs in several aspects. First, by embedding an SPE into a photonic cavity or resonator, the photon radiative rate can be increased through the Purcell effect, or a reduced pump power is needed to reach a similar photon emission rate. Second, photonic structures can route the emitted photons into a desired free space mode and increase the collection efficiency of the photons into an objective or fiber. Third, photonic structures can add an additional degree of freedom to the performance of the SPE, including, for instance, altering the spectrum, polarization, direction, and OAM of the photons. These controls can increase the degree of indistinguishability of photons from different trials. For epitaxial semiconductor QDs, it is possible to generate entangled photon pairs through the biexciton-exciton (XX-X) cascaded radiative process~\cite{PhysRevLett.96.130501}. In this case, a broadband photonic structure can increase the emission rate of both photons with close wavelengths and fidelity of entanglement.

\begin{figure}
  \includegraphics[width=\linewidth]{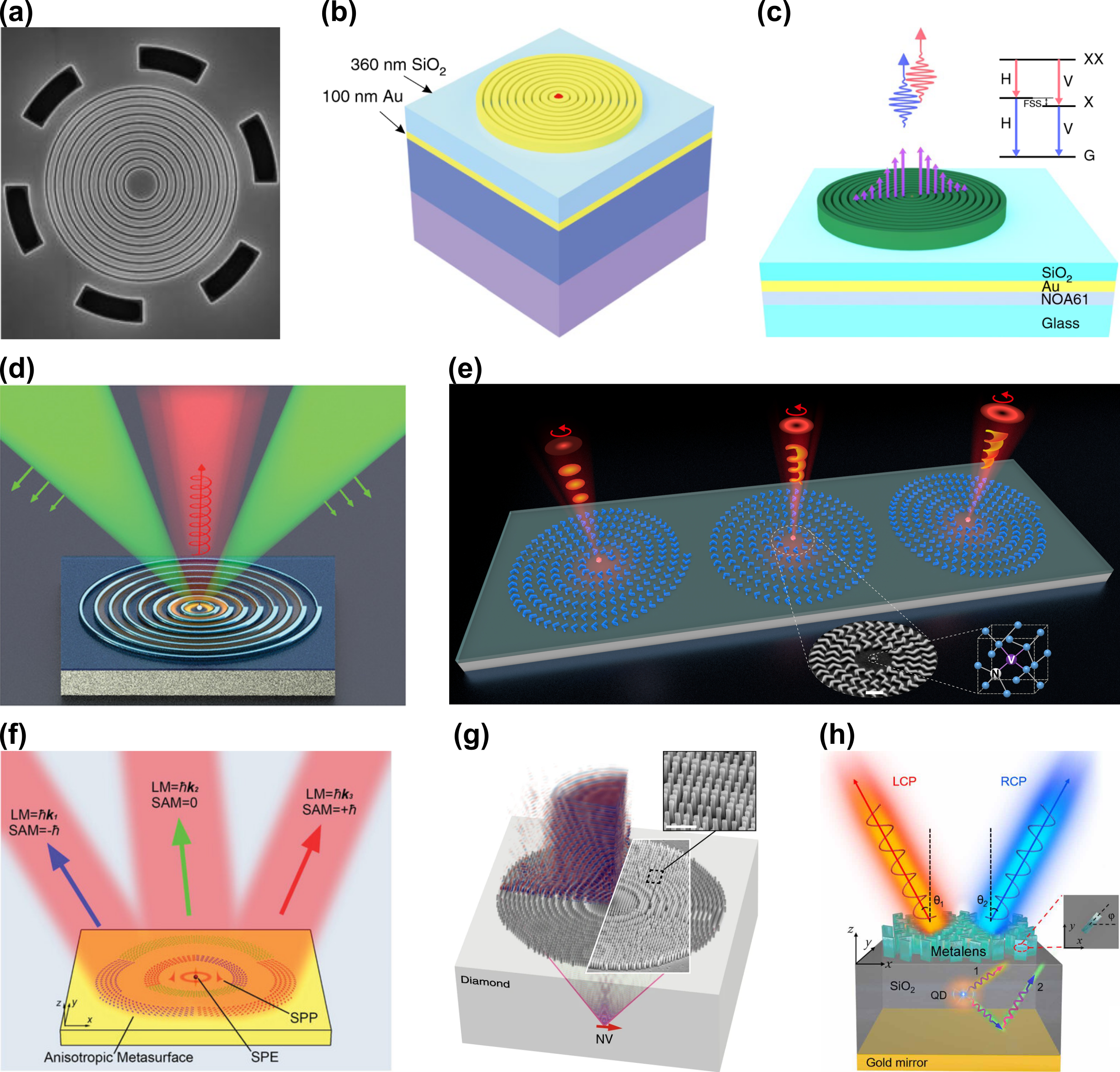}
  \caption{\textbf{Flat-optics SPE source with QDs and color centres in 3D materials.} 
  SPEs modulated by flat-optics nanostructures including (a,b,c,d) circular Bragg gratings (CBGs), (e,f) metasurfaces, and (g,h) metalenses.
  a)~CBG in a suspended GaAs membrane to enhance the emission efficiency of a single quantum dot located in the center through the Purcell effect and shape the emission mode into a nearly Gaussian profile which increases the collecting efficiency~\cite{davanco2011CircularDielectric}.
  b)~CBG with slight asymmetry to enable polarized single photon emission with efficiency exceeding the 50\% limit imposed by the polarization filtering~\cite{wang2019OptimalSinglephoton}.
  c)~CBG with broadband response to support generation of strongly entangled photon pairs with high brightness, degree of indistinguishability, and collecting efficiency~\cite{liu2019SolidstateSource}.
  d)~CBG with gradient width along the azimuthal direction to enable generation of circularly polarized single photons~\cite{kan2020MetasurfaceEnabledGeneration}.
  e)~Metasurface enabled the generation of circularly polarized single photons with orbital angular momentum~\cite{liu2023OnchipGeneration}.
  f)~Anisotropic metasurface enabled multichannel single photon emissions with on-demand spin states in each spatial channel~\cite{jia2023MultichannelSinglePhoton}.
  g)~Metalens to collimate the single photons emitted from a color centre inside the diamond~\cite{huang2019MonolithicImmersion}.
  h)~Metalens to collimate the single photons emitted from a QD and split the photons with different spin states into two directions~\cite{bao2020OndemandSpinstate}.
 Copyright Year, Publisher.
  }
  \label{fig:QD}
\end{figure}

Compared with other photonic structures, flat optical structures such as metasurfaces have advantages in the fabrication and precise positioning of SPEs. Among all flat optical structures, circularly symmetric or near-symmetric arrangements of nanostructures have received the most attention, including circular Bragg grating (bullseye grating), metalens, and circularly arranged metasurfaces. These structures are relatively easy to design as structure variation only happens along the radial direction. By putting an SPE in the center of the circular structure, it is straightforward to control the Purcell factor, emission pattern, polarization, and other properties of the photons. Circular structures are also most suitable for working with collecting objectives and fibers, which are all circular optics. For example, a circular Bragg grating (CBG) fabricated in a suspended GaAs membrane (Fig.~\ref{fig:QD}a) can shape the emission pattern of a single InAs quantum dot located in the center into a nearly Gaussian far field transverse profile, which facilitates an increased collection efficiency of a low-NA objective by 20 times when compared with a quantum dot in unpatterned GaAs~\cite{davanco2011CircularDielectric}. The same CBG also accelerates the spontaneous emission rate of the emitter through Purcell enhancement. An increase in the emission rate will minimize the dephasing effect and thus augment the coherence of the SPE emission. This can increase the indistinguishability of the emitted photons. Particularly, employing the resonance or cavity effect can potentially mitigate the disparity in coherence lengths between the coherent metasurface and incoherent photon emission, thereby enhancing the feasibility of utilizing metasurfaces for controlling SPE emission.
A similar concept has been demonstrated for epitaxial semiconductor QDs~\cite{sapienza2015NanoscaleOptical,moczala-dusanowska2020StrainTunableSinglePhoton,xia2021EnhancedSinglePhoton,kolatschek2021BrightPurcell,xu2022BrightSinglephoton,jeon2022PlugandPlaySinglePhoton,barbiero2022HighPerformanceSinglePhoton,li2023BrightSemiconductor,xiong2023EfficientGeneration,nawrath2023BrightSource} and color centres in diamond~\cite{kan2020MetasurfaceEnabledGeneration, li2015EfficientPhoton, komisar2023MultipleChannelling}. Importantly, by introducing a small asymmetry to the CBG (Fig.~\ref{fig:QD}b), it is possible to break the 50\% efficiency limit imposed by polarization filtering. Based on this idea, Hui Wang and co-workers reported a QD single photon source emitting polarized single photons with an efficiency of 56\% and an indistinguishability near 100\%, which paves a way to realize an ideal polarized single photon source~\cite{wang2019OptimalSinglephoton}. Furthermore, a broadband CBG (Fig.~\ref{fig:QD}c) is also able to improve the performance of generating entangled photon pairs from QDs~\cite{liu2019SolidstateSource, wang2019OnDemandSemiconductor, rota2022SourceEntangled}. A gradient-width bullseye grating has enabled the direct generation of circularly polarized single photons from a nitrogen-vacancy center in a nanodiamond (Fig.~\ref{fig:QD}d)~\cite{kan2020MetasurfaceEnabledGeneration}. 

Metasurfaces comprised of separate nanostructures with various at-will shapes have more design degrees of freedom when compared with CBG. This enables more complex control of the SPEs through, for example, simultaneous control of multiple degrees of freedom of the single photons. Examples include single photon emissions with specific orbital angular momentum while having desired linear~\cite{liu2023UltracompactSinglePhoton} or circular polarizations (Fig.~\ref{fig:QD}e)~\cite{liu2023OnchipGeneration}, and multichannel generation of single photons with different spin angular momenta (Fig.~\ref{fig:QD}f)~\cite{jia2023MultichannelSinglePhoton}.

When the SPE is not on the front surface where nanostructures can be made, a metalens is a promising flat optical element to efficiently control the emitted photons from the SPE.  
% A metalens is another flat optical element that can be patterned into flat optical structures to efficiently collimate the emitted photons from an SPE when the emitter is not on the front surface. 
This is done by designing the metalens such that the SPE is at the focal point. As shown in Fig.~\ref{fig:QD}g, a metalens was fabricated on the surface of the diamond to collimate the emission from a nitrogen-vacancy center inside the diamond~\cite{huang2019MonolithicImmersion}. Apart from collimation, metalenses can further manipulate the spin state and direction of the single photons from a semiconductor QD (Fig.~\ref{fig:QD}h)~\cite{bao2020OndemandSpinstate}. Recently, a metalens was also directly fabricated on the wide-bandgap silicon carbide~\cite{schaeper2022MonolithicSilicon}, and integration with SPEs hosted within SiC is a promising avenue of research.

\subsection{2D materials}

Initial efforts to isolate SPEs in transition metal dichalcogenides (TMDs) like WSe$_2$ relied on local excitons bound to intrinsic atomic defects present in as-grown~\cite{He:2015-497:NNANO} or exfoliated~\cite{Koperski:2015-503:NNANO} monolayers. The confining potential provided by the defects yielded single photon emission with bright, narrow photoluminescence (PL) peaks slightly redshifted from the broader delocalized monolayer exciton at cryogenic temperature, although the localized emission was randomly distributed across each TMD flake. An effective step towards the systematic investigation of these single photon sources was provided by Palacios-Berraquero et al.~\cite{Palacios-Berraquero:2017-15093:NCOM}, who used arrays of nanopillars patterned onto a silica substrate to artificially impose confining potentials in TMD monolayers through local strain gradients. Exfoliated monolayer WSe$_2$ was dry-transferred onto the patterned substrate, draping tightly over the nanopillars, with bright and localized cryogenic PL observed over unpierced regions of the strained monolayer, as shown in Fig.~\ref{fig:2D}(a). 
% In the best case, this method produced single photon emitters at around one from every three pillar sites, accounting for pillars which pieced the monolayer or sites which hosted two or more quantum emitters.  While the single emitter probability and the spectral stability were found to improve with increasing nanopillar height, no correlation was observed between nanopillar height and emission peak wavelength, with PL peaks uniformly distributed from 730 nm to 820 nm. Similar results were reported when using WS$_2$, and when using drop cast nanodiamonds in place of the fabricated nanopillars. 

\begin{figure}
  \includegraphics[width=\linewidth]{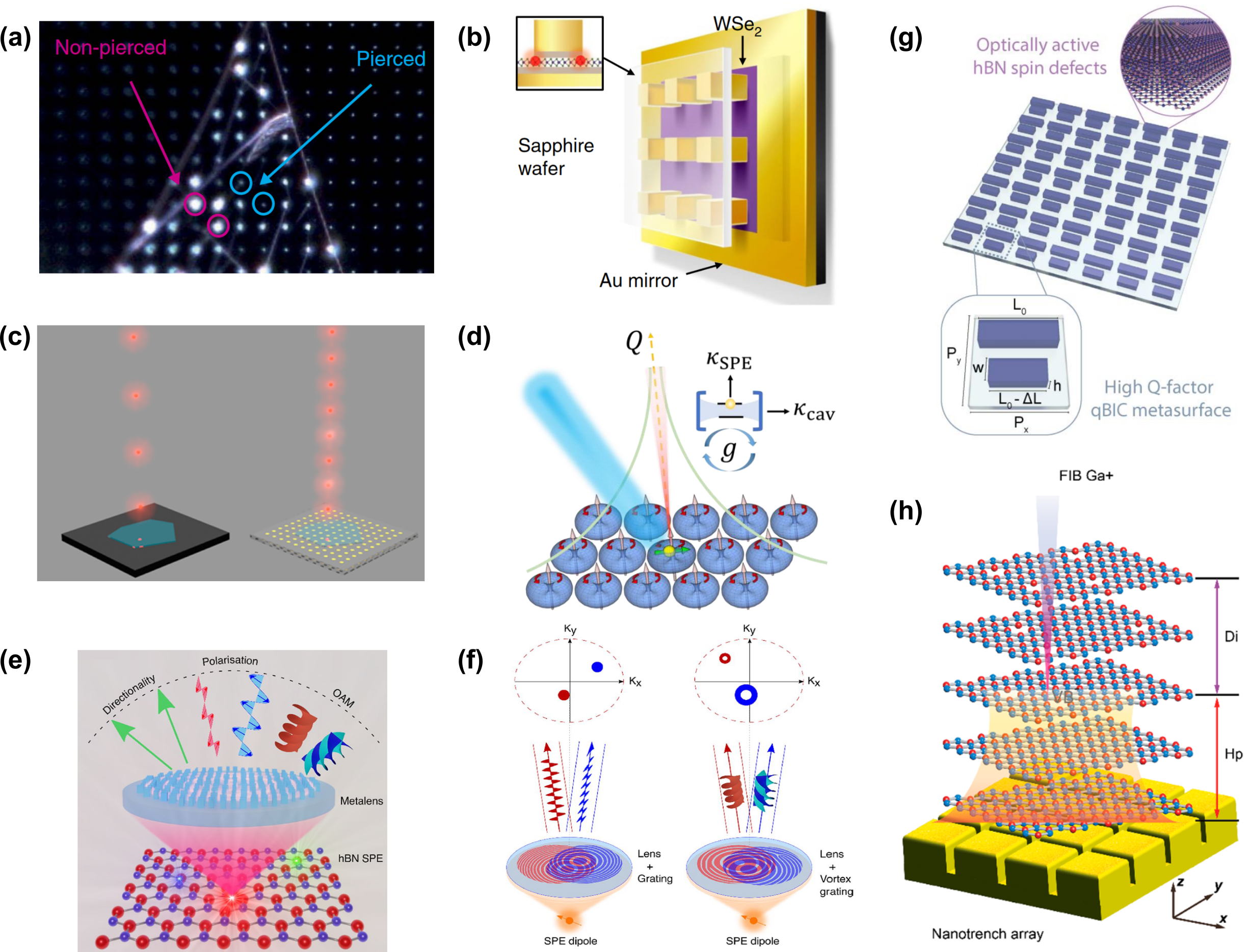}
  \caption{\textbf{Single photon emission from 2D material flat optics.}  a)~Spatial control of SPE creation in TMDs using localized strain~\cite{Palacios-Berraquero:2017-15093:NCOM}. b)~Closed plasmonic nanocavities leading to enhanced emission from SPEs in TMDs~\cite{Luo:2018-1137:NNANO}.
  c)~Enhancement of SPE emission in hBN using an open plasmonic nanocavity array~\cite{Tran:2017-2634:NANL}. d)~Strong coupling between a metasurface BIC mode and monolithically integrated hBN SPEs~\cite{do2022RoomTemperature}. e-f)~Controlling the photon degrees of freedom of SPEs in hBN using a multifunctional hBN metalens~\cite{liArbitrarily2023s}.  
  g) Quantum efficiency enhancement of spin-active defects in hBN using a metasurface qBIC mode~\cite{Sortino:2306.05735:ARXIV}. h) PL enhancement of spin defects in hBN using plasmonic resonances in a trenched metasurface~\cite{caiSpin2023}. Copyright Year, Publisher.
}
  \label{fig:2D}
\end{figure}

Beyond simply activating the quantum emitters, a metasurface can be designed to simultaneously introduce local strain in the integrated 2D material and act as a cavity for electric field enhancement, leading to modulation of the recombination rate. Luo et al.~\cite{Luo:2018-1137:NNANO} used an array of gold nanocubes on a sapphire substrate to support monolayer WSe$_2$, finding strain-activated single photon emission over each corner of the top face of each nanocube with high probability. An initial characterization of these emitters through power saturation and exciton lifetime measurements provided a baseline for the subsequent analysis of the cavity effects. A closed plasmonic gap mode was then introduced by transferring the entire WSe$_2$/Au/Sapp structure onto a gold mirror, as illustrated schematically in Fig.~\ref{fig:2D}(b), and the performance of the same emitters was reassessed. The plasmonic resonance of the gold nanocavity exhibited strong spectral overlap with the WSe$_2$ PL emission, leading to significant Purcell enhancement and an average 15-fold reduction in exciton lifetime. The gap width in this approach is limited only by the thickness of the TMD monolayer together with the 2 nm aluminium oxide spacer layer required to isolate the gold components, which produces a very small mode volume. However, the plasmonic resonance was broad in order to achieve spectral overlap with the wide distribution of exciton emission wavelengths, leading to a small Q factor. Hence, spatial control of emitter creation should ideally be accompanied by spectral control for optimal Purcell enhancement. 

In the TMD system, the excitonic nature of the SPE emission necessitates cryogenic conditions. In contrast, the wide bandgap of hBN allows for deeply protected defect states that provide SPEs in the visible wavelength range at room temperature~\cite{stewart2021QuantumEmitter, fournierPositioncontrolled2021, hoese2020MechanicalDecoupling, liu2023SinglePhoton, ziegler2019DeterministicQuantum, patel2022ProbingOptical, tan2022DonorAcceptor}. 
% The first reports of quantum emission from many-layer flakes of hBN employed high temperature annealing~\cite{tranQuantum2016, jungwirthTemperature2016} or sparse electron beam irradiation~\cite{tranRobust2016, exarhosOptical2017} to activate emitters randomly across the sample. Similar results were achieved through oxygen plasma etching~\cite{voglFabrication2018} and UV ozone treatment~\cite{liPurification2019}, and UV emission was observed in cathodoluminescence measurements~\cite{bourrellierBright2016}. 
Room temperature PL spectra typically feature a bright and narrow zero phonon line (ZPL) together with a dimmer phonon side band (PSB) redshifted by around 160 meV. The ZPLs occur over a wide range, from 550 nm to 730 nm, 
% correlations between PSB shape, the presence of metastable shelving states evident in the second order correlation function, and the alignment of absorption and emission dipoles indicate 
and distinct classes of emitters were proposed~\cite{exarhosOptical2017}. The exact defect structures are yet unknown, however, carbon-related defects are a likely candidate for the visible emitters~\cite{mendelsonIdentifying2021}.  

% Metasurfaces
Nanoscale pillars have been used to control the spatial distribution of visible SPEs in hBN. Few-layer flakes draped over an array of flat-topped silica nanopillars, similar to the scheme presented in Fig.~\ref{fig:2D}(a), yielded strain-induced SPEs with a normally distributed range of ZPLs centered at 570 nm~\cite{prosciaNeardeterministic2018}. An alternative approach utilized direct chemical vapor deposition (CVD) growth onto the nanopillars, with the number of SPEs per nanopillar found to be proportional to nanopillar diameter~\cite{liScalable2021}. The smallest investigated diameter of 250 nm resulted in a near unity yield of SPEs, with ZPLs clustered around 600 nm, which was attributed to the reduction in available CVD nucleation sites and hence defect creation over the smaller pillar surface area. 
% It should be noted that visible SPEs can also be positioned using opaque masks~\cite{stewartQuantum2021} or femtosecond laser pulses~\cite{ganLargeScale2022}, with comparable precision and ZPL distribution. In fact, the greatest control over both emitter spatial location and ZPL wavelength uses brief electron beam irradiation to produce defects with quantum emission at 436 nm~\cite{shevitskiBluelightemitting2019, fournierPositioncontrolled2021, galeSiteSpecific2022}, although this shorter wavelength emission is not desirable for coupling to dielectric or plasmonic metasurfaces. 

Without spatial control of emitter activation, alignment with metasurface cavities is challenging. Tran et al.~\cite{Tran:2017-2634:NANL} were able to utilize the pick-and-place capabilities of hBN to overcome this issue. The left image in Fig.~\ref{fig:2D}(c) illustrates the initial PL characterization of visible emitters created via thermal annealing in a flake exfoliated onto a silicon substrate. The position of bright, single candidate emitters was recorded relative to lithographic markings, and the same flake was then transferred using a spin-coated PMMA layer onto a metasurface consisting of either gold or silver nanocube arrays, as shown in the right of Fig.~\ref{fig:2D}(c). Comparing the lifetime measurement of an emitter over the bare silicon substrate, with the same relocated emitter positioned over a nanocube, a modest Purcell enhancement was observed, leading to a doubling of brightness. The silver metasurface performed better, owing to the lattice plasmon resonances being more robust against lossy interband transitions around the target ZPL wavelength. Further enhancement may be possible by closing the gap mode, as depicted in Fig.~\ref{fig:2D}(b), or by scaling the nanocube dimensions to optimize cavity spectral overlap after the identification of emitter ZPLs. Purcell enhancement in the weak coupling regime can also be achieved with dielectric nanostructures, where cavity resonances have been tuned in situ using, for example, piezoelectric actuators to set the spacing in a dielectric Bragg mirror~\cite{voglCompact2019a} or gas condensation to modulate the geometry of a 1D photonic crystal cavity~\cite{frochPurcell2022}. Coupling emission to guided modes in dielectric nanostructures offers additional capabilities for metasurface integration~\cite{prosciaMicrocavity2020, liIntegration2021a}.  

The use of sub-diffractive arrays in metasurface design allows for the creation of BICs, optical modes with relatively low dissipative decay rates. A BIC mode could be a route toward strong electron-photon coupling in conditions where a large mode volume is unavoidable. This approach was pursued by Do et al.~\cite{do2022RoomTemperature}, using a thin hBN film transferred onto an array of TiO$_2$ nanopillars as shown in Fig.~\ref{fig:2D}(d). The BIC resonance is associated with a magnetic dipole along each nanopillar vertical axis, with a corresponding circulating electric field throughout each nanopillar which can couple to a well-aligned emitter dipole. Energy-resolved back focal plane imaging revealed weak coupling between an emitter with ZPL slightly detuned from the BIC resonance, and evidence suggestive of strong coupling for an emitter at zero detuning. The BIC resonance energy can be selected by defining the nanopillar diameter and array spacing, however, this approach is still subject to the probabilistic spectral and spatial location of hBN emitters. 

% Beyond their use in generating quantum light from 2D materials, metasurfaces are ideal for manipulating the properties of single photons in flight. These properties include propagation direction, polarization, and OAM, as depicted in Fig.~\ref{fig:2D}(e). 
One of the unique properties of SPEs in hBN is their brightness. This enabled an interesting demonstration that combined the control of propagation direction, polarization, and OAM of photons emitted from SPEs employing a metasurface, as depicted in Fig.~\ref{fig:2D}(e).
% Interestingly, these functionalities can even be coupled together in a single metasurface. 
Li et al.~\cite{liArbitrarily2023s} fabricated a metalens consisting of an array of hydrogenated amorphous silicon meta-atoms of varying lengths and widths. When the dimensions vary linearly across the metalens, transmitted light is deflected according to polarization, while a spiral variation is employed to structure the OAM of the incident beam. Superimposing these lens geometries yields a polarizing beam splitter with output modes possessing distinct topological charges (Fig.~\ref{fig:2D}(f)).  

% Birefringence in hBN emerges from negative isotropic in-plane permittivity and positive out-of-plane permittivity, resulting in an intrinsic isotropic dispersion of surface phonon polaritons (SPhP). This would manifest as circular SPhP wavefronts propagating from an out-of-plane dipole source. But total anisotropic permittivity can be achieved by introducing a linear grating in the hBN, which results in a hyperbolic in-plane SPhP dispersion that inhibits propagation of polaritons along the direction perpendicular to the grating. Such a structure supports distinctive concave wavefronts, as shown in Fig.~\ref{fig:2D}(f). Li et al.~\cite{liInfrared2018} developed a mid-wavelength IR hyperbolic metasurface (HMS) by etching a deeply subwavelength grating in hBN using electron beam lithography. Polaritons launched from a gold nanoantenna produced a concave interference fringe pattern, as revealed by near-field scanning optical microscopy. The dispersion is sensitive to grating width and probe wavelength, with additional variation expected for each 2D material as well as heterostructures and Moire superlattices. 

% The exact defect structure of single-photon emitters in hBN remains elusive, however, emission from ensembles of the boron vacancy (V$_B^-$) defect has been well studied
An important defect within the hBN family is the negatively charged boron vacancy (V$_B^-$)~\cite{gottschollRoom2021}. The spin of the V$_B^-$ defect is optically addressable at room temperature via the optically detected magnetic resonance (ODMR) that enables precise measurements of magnetic and electric fields~\cite{kumar2022MagneticImaging, healeyQuantum2023} at the nanoscale. One of the major drawbacks of V$_B^-$
% compared to more mature spin systems in diamond, is the relatively low optically detected magnetic resonance (ODMR) contrast, which limits the field measurement sensitivity. 
is its low quantum efficiency that limits the signal-to-noise ratio in ODMR measurements and requires longer integration times.
Metasurfaces are an avenue for enhancing the intrinsically low quantum efficiency and weak PL intensity of the V$_B^-$ defect. Sortino et al.~\cite{Sortino:2306.05735:ARXIV} introduced an asymmetric scaling into an hBN nanopattern in order to access a high-Q qBIC resonance, as shown in Fig.~\ref{fig:2D}(g). The resonance energy was easily specified by controlling the magnitude of the asymmetry parameter $\Delta L$, providing flexibility in the spectral overlap with the naturally broad V$_B^-$ emission. Nitrogen beam irradiation introduced V$_B^-$ ensembles directly into the hBN metasurface, and such monolithic integration leads to greater coupling efficiency than is possible with hybrid structures using multiple materials. While the reduction in lifetime induced by this qBIC metasurface was small, ensemble brightness was found to increase around 20-fold, leading to a stronger ODMR signal. 

The case of hybrid coupling was explored with nanocubes~\cite{mendelson2022CouplingSpin, xu2023GreatlyEnhanced} as well as trenched structures, where Cai et al.~\cite{caiSpin2023} found that the plasmonic resonance of a gold metasurface was redshifted with every extra layer of hBN transferred on top. As Fig.~\ref{fig:2D}(h) shows, the field enhancement provided by the gap plasmon mode drops off in proportion to the surface normal distance Hp, whereas the V$_B^-$ ensemble depth Di is ideally minimized for the greatest magnetic field sensitivity. A combination of nanotrench spacing and hBN depth contributed to the optimal coupling conditions, which yielded a 40-fold increase in brightness compared to V$_B^-$ ensembles located over flat regions of the gold substrate. Hence, metasurface performance appears to depend closely on the nature of the integration with the target material, and this consideration should be incorporated at the design stage. 

The flat optics devices utilizing SPEs in hBN~\cite{liArbitrarily2023s, spencer2023MonolithicIntegration} have indeed demonstrated purities and, where applicable, indistinguishabilities comparable to the same class of emitter in bulk, unstructured material~\cite{shields2007SemiconductorQuantum, kurtsiefer2000StableSolid, martnez2016EfficientSingle}. This indicates that increases in brightness and enhanced polarization control can hence be achieved with minimal drawbacks.

\section{Challenges and future prospects}
% \subsection{Comparison and challenges of flat-optics quantum light sources}
% Achieving highly efficient, coherent, deterministic, scalable, and room-temperature quantum light sources is a paramount objective. This achievement is an essential step towards practical applications of quantum technologies. Two distinct pathways, utilizing photon pairs and single photons, present unique advantages and challenges in pursuit of this goal. 
The combination of flat optics and quantum light sources is rapidly gaining momentum as a research field in its own right. Achieving highly efficient, coherent, and deterministic quantum light sources is a paramount objective, with room temperature operation being an added benefit. However, where metasurfaces add a critical advantage is with the possibilities of functionalities. As was described above, metasurfaces can enable wavefront control or spatial localization with a single purposely designed nanostructure. It is important to emphasize that the metasurfaces as a whole are agnostic to the use of probabilistic (SPDCs) or deterministic sources. Both avenues have their own promises and should be pursued in the coming years.

The SPDC process inherently allows the generation of indistinguishable photons, facilitating the preparation of complex quantum entanglement. 
% Notably, this process operates at room temperature, enhancing its suitability for most of quantum applications. 
There are two primary challenges in the generation of photon pair sources via SPDC. The first one is the strict requirements for longitudinal phase matching conditions, which need dedicated temperature control. The incorporation of flat optics has successfully addressed this concern and provides a potential for scalable production of photon pair sources with added functionalities. However, this solution comes at the cost of a significantly reduced photon pair rate due to the ultra-short interaction length of the materials. Novel approaches (such as triple resonances at pump, signal and idler photon wavelength, improved design and fabrication of metasurfaces for high-Q resonances, etc.) ought to be developed to improve the rate of photon pairs from nonlinear metasurfaces. The second challenge arises from the probabilistic nature of the SPDC process. The chances of a pump photon splitting into signal and idler photons are rather low. While the photon-pair rate in the state-of-the-art experiments has reached the megahertz range, the generation remains spontaneous and non-deterministic, posing difficulties for quantum computing applications. The potential of resolving this issue through the utilization of ultrafast switching and photon routing in metasurfaces remains an unexplored avenue.

The production of photon pairs via SPDC was recently reported from 2D materials. %NbOCl$_2$~\cite{Guo:2023-53:NAT}. 
%Such SPDC sources, 
Due to their remarkable thinness and high efficiency, they provide an avenue for developing miniaturized on-chip SPDC sources and high-performance photon modulators. We anticipate that their potential extends beyond unstructured 2D materials. Metasurfaces composed of 2D material structures offer unlimited potential towards the next-stage development of quantum light sources.

Solid-state SPEs and QDs, in contrast, offer high efficiency and deterministic characteristics, enabling precise control over the timing of emitted photons. This property is indispensable for applications such as quantum repeaters where precise synchronization and manipulation of single photons are required. Although the development of metasurface-based quantum emitters has led to Purcell enhancement, improved collection efficiency, and complete control of light over its emission direction and polarization, it is still an open question to realize coherent and indistinguishable single photons. Addressing this challenge may be possible by employing metasurfaces with ultra-high Q factors. Furthermore, single photon emission from 2D materials primarily occurs within the visible wavelength range. The development of infrared SPEs will be a critical step towards the applications of quantum telecommunications. In this regard, the exploration of tunable metasurfaces in the generation of quantum light sources holds significant promise to dynamically control quantum states and will bring this field to a new milestone. Additionally, combining multi-functional metasurfaces with integrated platforms such as SPEs in silicon nitride waveguide~\cite{Roomtemperature2021, senichevSilicon2022} is another interesting direction worthy of further exploration.

The ability to control photon polarization and angular momentum using metasurfaces could yield compact in situ filtering capabilities. Whether SPEs are excited with above-resonant or on-resonant laser energy, scattered laser light must be filtered to isolate the quantum light. In particular, on-resonant excitation is preferable for the greatest photon coherence, however, the separation of the laser from the ZPL photons requires complicated bulk cross-polarization optics or emission into orthogonal waveguide modes. Ideally, a laser extinction method would be integrated with a metasurface hosting the SPE, which would allow for the generation of maximally coherent single photons on demand. 

A key promise with metasurfaces and defects may in fact arise from the area of quantum sensing. Integration of spin defects with a metasurface on top of an optical fibre may be utilized to engineer a plug-and-play quantum sensing device. Fabrication of fibre-based metasurfaces is an emerging field~\cite{zhao2022optical, principe2017optical}, and the ability to deterministically transfer 2D materials onto fibres may play a key role in this direction.

An interesting avenue that is still in its infancy is nonlocal metasurfaces for SPEs.
%~\cite{Shastri:2023-36:NPHOT, Song:2021-1224:NNANO}. 
Nonlocal resonances feature strong angular dispersion, a characteristic not present in local metasurfaces. Exploiting this unique property provides the opportunity to tailor the emission directions and photon frequency of quantum light sources. Additionally, its unique advantages in wavefront control have not yet been implemented in quantum light sources.

Finally, incorporating inverse design~\cite{li2022empowering, pestourie2018inverse, elsawyNumerical2020a, moleskyInverse2018a, maDeep2021,jiRecent2023}, including topology optimization, evolutionary optimization, and machine learning techniques, has arisen as a revolutionary approach for addressing the challenges associated with metasurface design. Unlike conventional forward design methods, inverse design is characterized by its method-agnostic nature and its objective of solving physics problems through the use of mathematical tools. Notably, inverse design remains unexplored in the context of flat optics for quantum light sources, representing a promising avenue to unlock novel possibilities for nanoscale control over light sources with multifunctional capabilities.

\medskip
\textbf{Acknowledgements} \par %delete if not applicable))
The authors acknowledge financial support from the Australian Research Council (CE200100010 and FT220100053) and the Office of Naval Research Global (N62909-22-1-2028).

% References
\medskip

% Use the following code if you wish to generate your bibliography with BibTeX;
% replace the string "MSP-template" below with the name(s) of
% the BibTeX data base(s) you want to use.
% The resulting bibliography-output (the content of the .bbl file)
% must be pasted back into this file before submission.
% Please also include your BibTeX data base file(s) in your submission
% so that we can re-run BibTeX if necessary.
%
\bibliographystyle{MSP}
\bibliography{review_SPDC,review_SQE,review}

\end{document}